\documentclass[%
onecolumn,pra,
superscriptaddress,
 amsmath,amssymb,
 aps,
11pt,
]{revtex4-1}

\usepackage{graphicx}
\usepackage{array}
\usepackage{dcolumn}
\usepackage{bm}
\usepackage{listings}
\usepackage{scalefnt}
\usepackage{xcolor}
\usepackage{multirow}
\usepackage{siunitx}
\usepackage{booktabs}
\usepackage{amsthm,amsfonts}
\usepackage{adjustbox}
\usepackage{tabularx}
\usepackage{lmodern}

\usepackage{braket}

\def\sz#1{\hat{\sigma}_z^{(#1)}}
\def\sx#1{\hat{\sigma}_x^{(#1)}}
\def\gb{\mathbf{\gamma},\mathbf{\beta}}
\def\W#1#2{\hat{W}_{#1}^{#2}}
\def\Wd#1#2{\hat{W}_{#1}^{#2\,\dagger}}

\def\sC#1{\hat{\vec{\sigma}}^{(C)}_{#1}}

\def\sG#1{\hat{\vec{\sigma}}^{(G)}_{#1}}

\DeclareMathOperator*{\dprime}{\prime \prime}

\renewcommand{\vec}[1]{\boldsymbol{#1}}

\begin{document}

\title{Practical optimization for hybrid quantum-classical algorithms}

\author{Gian Giacomo Guerreschi}
\email{Email: gian.giacomo.guerreschi@intel.com}
\affiliation{Intel Labs, Santa Clara, California 95054, United States}
\author{Mikhail Smelyanskiy}
\affiliation{Intel Labs, Santa Clara, California 95054, United States}
\date{\today}

\begin{abstract}
A novel class of hybrid quantum-classical algorithms based on the variational approach have recently emerged from separate proposals addressing, for example, quantum chemistry and combinatorial problems. These algorithms provide an approximate solution to the problem at hand by encoding it in the state of a quantum computer. The operations used to prepare the state are not a priori fixed but, quite the opposite, are subjected to a classical optimization procedure that modifies the quantum gates and improves the quality of the approximate solution.
While the quantum hardware determines the size of the problem and what states are achievable (limited, respectively, by the number of qubits and by the kind and number of possible quantum gates), it is the classical optimization procedure that determines the way in which the quantum states are explored and whether the best available solution is actually reached.
In addition, the quantities required in the optimization, for example the objective function itself, have to be estimated with finite precision in any experimental implementation. While it is desirable to have very precise estimates, this comes at the cost of repeating the state preparation and measurement multiple times. Here we analyze the competing requirements of high precision and low number of repetitions and study how the overall performance of the variational algorithm is affected by the precision level and the choice of the optimization method. Finally, this study introduces quasi-Newton optimization methods in the general context of hybrid variational algorithms and presents quantitative results for the Quantum Approximate Optimization Algorithm.

\end{abstract}

\maketitle


\vspace{5mm}
\section{Introduction}

After decades of slow but steady progress, the technology behind quantum computation is undergoing an impressive development in recent years. Systems that were once confined to small proof-of-concept implementations in academic laboratories are growing into quantum devices eager to challenge the most powerful supercomputers. Two factors fundamentally contributed to this change of perspective: First, a few hardware platforms have reached the level of control required to scale-up the number of components, \emph{i.e.} of qubits, while at the same time being able to apply longer sequences of high-fidelity quantum gates \cite{Riste2015,Barends2016,Monz2016,Debnath2016,Carolan2015,Mower2015,Veldhorst2015}.
Second, novel quantum algorithms have been proposed that try to deal with decoherence and noise without explicitly including quantum error correction protocols and, thus, avoiding their massive overhead. To this category belong some of the latest schemes proposed to achieve quantum supremacy, like boson sampling \cite{Aaronson2011} and random quantum circuits \cite{Boixo2016}, but also hybrid quantum-classical algorithms based on the variational approach \cite{Peruzzo2014,McClean2016,Farhi2014,Farhi2016a}.

A representative example of these hybrid schemes is the Variational Quantum Eigensolver (VQE) algorithm recently proposed to find approximate solutions to the ground state energy problem of quantum Hamiltonians \cite{Peruzzo2014} like, but not exclusively, those Hamiltonians used to describe electrons in molecules.
Around the same time, the Quantum Approximate Optimization Algorithm (QAOA) was introduced to solve constraint satisfaction problems of binary variables \cite{Farhi2014}, especially those for which no classical efficient algorithm is available.
The idea is as follows: Variable assignments can be mapped into qubit states in the register of a quantum computer and their properties, among which the number of satisfied clauses, computed by measuring suitable observables and averaging the outcomes. Modifying the qubit state corresponds to exploring different superpositions of assignments and, if done following a systematic approach, leads to states with the desired property. In the case of QAOA, one is typically interested to identify an assignment that satisfied as many constraints as possible and the result of the algorithm is an approximated solution based, first, on what states can be prepared by the quantum circuit and, second, on how well one is able to identify the best performing state among them.

As explicitly stated in the original work \cite{Peruzzo2014}, at the origin of VQE there was the idea of a ``device dependent ansatz'', meaning that the set of available states was exactly formed by all the state reachable by the limited capabilities of specialized, small quantum systems. Researchers recognized that such approach was unsatisfactory when implemented on larger devices capable of many more quantum gates leading to a dramatic growth of the number of parameters: In fact the exploration of the parameter space quickly became the new bottleneck. In the context of molecular energies, this limitation was overcome by using established ``theoretical ansatz'' (like unitary coupled-cluster states \cite{McClean2016,OMalley2016}) described by a manageable (\emph{i.e.} polynomial) number of parameters. In QAOA the characterization of the quantum gates was carried to the extreme and only two operations, alternated and repeated in sequence, were necessary.

Irrespective of the specific ``ansatz'' used to parametrize the final state, one needs a way to select promising sequences of gates that improve the quality of the solution while exploring as few options as possible. The usual approach implements methods that use the last few values of the objective function, corresponding to the performance of the last few states, to predict new candidates \cite{Nelder1965,Powell1964,Kelley1999,McClean2016}. These schemes are called ``derivative free'' since they do not take advantage of the smoothness of the objective function to access and exploit its gradient. Here we discuss how to apply more advanced optimization methods that depend on the evaluation of the gradient, obtained either through its analytical expression or following a finite-difference approach.

To set the comparison between optimization methods on a firm ground that reflects the actual cost of experimental implementations, we do not simply consider how many times the objective function and its gradient have to be evaluated, but count the number of repetitions necessary to estimate each quantity. With the term single repetition we mean the complete state preparation and the measurement of all, or a part of, the qubit register. The estimate of the expectation value of a single observable requires several repetitions, and their number increase quadratically with the precision required. However, while lowering the precision reduces the repetition count, it also risks to deteriorate the effectiveness of the overall optimization process.

Here we quantitatively describe the impact of finite precision on the repetition cost and compare the performance of different optimization methods for various levels of precision. The numerical simulations are performed for a specific hybrid algorithm addressing combinatorial problems, namely the Quantum Approximate Optimization Algorithm (QAOA) \cite{Farhi2014,Farhi2016a}.

\vspace{5mm}
\section{Hybrid quantum-classical algorithms}
\label{sec:hybrid_algorithms}

\begin{figure*}[t!]
\begin{center}
\includegraphics[width=0.8\linewidth]{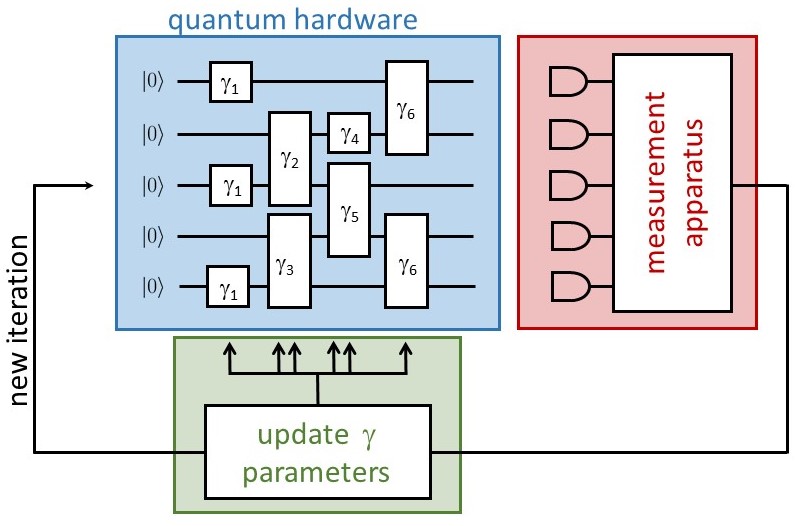}
\caption{Illustration of the three common steps of hybrid quantum-classical algorithms. These steps have to be repeated until convergence or when a sufficiently good quality of the solution is reached. 1) State preparation involving the quantum hardware capable of tunable gates characterized by parameters $\gamma_n$ (blue), 2) measurement of the quantum state and evaluation of the objective function (red), 3) iteration of the optimization method to determine promising changes in the state preparation (green). Notice that a single parameter $\gamma_n$ may characterize more than one gate, for example see $\gamma_1$ and $\gamma_6$ in the blue box. In practice, many state preparations and measurements are necessary before proceeding with a single update of the parameters.}\label{fig:hybrid_scheme}
\end{center}
\end{figure*}

The class of hybrid quantum-classical algorithms that is the subject of this work is characterized by three distinct steps. The first one corresponds to the state preparation and is achieved by applying a sequence of gates, described by the corresponding set of parameter values, on an initial reference state. The second step is the measurement operation in which one measures the quantum state, records the outcomes and analyze them to obtain the value of the objective function corresponding to the prepared state. The third step is the classical optimization iteration that, based on previous results, suggests new parameter values to improve the quality of the state. We pictorially illustrate these three parts and their interplay in Fig.~\ref{fig:hybrid_scheme}.

As mentioned, the goal of variational algorithms is to find an approximate solution to certain problems. The quality of such approximation is given by the value of the objective function that one desires to maximize (or minimize). The objective function is expressed as a quantum observable, noted here with $\hat{C}$, of the qubit register. It can be a genuinely quantum quantity, as is the case for the energy of molecular systems, or classical in nature, for example when it is associated to combinatorial optimization, scheduling problems or financial modeling. Given the quantum register in state $\ket{\phi}$, the objective function is given by the expectation value $\bra{\phi} \hat{C} \ket{\phi}$.

The choice of the state ansatz specifies how state $\ket{\phi}$ is prepared. In general, one applies a sequence of $p$ gates $\W{1}{p}(\mathbf{\gamma})=\W{1}{p}(\gamma_1,\gamma_2,\ldots,\gamma_p)$:
\begin{equation}
	\W{1}{p}(\mathbf{\gamma}) = \hat{U}_p(\gamma_p) \, \ldots \, \hat{U}_2(\gamma_2) \, \hat{U}_1(\gamma_1) \, ,
\end{equation}
to a reference state $\ket{\phi_0}$. Each gate is characterized by a parameter $\gamma_n$, and gates $\hat{U}_n(\gamma_n)$ and $\hat{U}_m(\gamma_m)$ do not have to commute for $n\neq m$.
The final state is:
\begin{equation}
\label{eq:final_state_gen}
    \ket{\mathbf{\gamma}} = \ket{\gamma_1,\ldots,\gamma_p} = \W{1}{p}(\mathbf{\gamma}) \ket{\phi_0} \, .
\end{equation}

For practical applications, it is helpful to express $\hat{C}$ as a linear combination of Hermitian operators that can be easily measured in experiments. In fact, it is always possible to express $\hat{C}=\sum_{\nu=1}^{k_C} c_\nu \, \sC{\nu}$ as the weighted sum of products of a few Pauli matrices and impose the coefficients $c_\nu$ to be real and positive (possibly including a minus sign in $\sC{\nu}$). More general choices of $\sC{\nu}$ are possible (\textit{i.e.} not only products of Pauli operators), since the constraints come from experimental limitations. Finally, notice that for most problems of interest, the number $k_C$ of terms is polynomial in the number of qubits $N$, typically a quadratic, cubic or quartic function. If the problem is classical, the observable $\hat{C}$ is diagonal in the computational basis and $\sC{\nu}$ pairwise commute (even better, in this case $\hat{C}$ can be measured directly). Staying general, the explicit form of the objective function is:
\begin{equation}
\label{eq:objective_function_gen}
	F_p(\mathbf{\gamma}) = \bra{\mathbf{\gamma}} \hat{C} \ket{\mathbf{\gamma}}
    					 = \sum_{\nu=1}^{k_C} c_\nu \, \bra{\mathbf{\gamma}} \sC{\nu} \ket{\mathbf{\gamma}} \, .
\end{equation}

Observe that a single measurement of $\ket{\mathbf{\gamma}}$ cannot directly provide the value $F_p(\mathbf{\gamma})$. Instead, it is computed by repeating the state preparation and measurement steps and accumulating enough outcome statistics to estimate the expectation value of $\hat{C}$.
We note with $F_{p,\epsilon}(\mathbf{\gamma})$ the estimator within precision $\epsilon$, meaning that it belongs to a stochastic distribution centered in $F_p(\mathbf{\gamma})$ and with standard deviation $\epsilon$.

Based on the frequentist approach to probabilities and considering each experimental repetition as independent, one can compute the number of repetitions $M$ to achieve such level of precision:
\begin{equation}
\label{eq:objective_function_cost}
	M \, \geq \, \frac{\mathrm{Var} [\hat{C}] }{\epsilon^2} \,
    					 = \, \frac{\sum_{\nu=1}^{k_C} c_\nu^2 \, \mathrm{Var} [\sC{\nu}]}{\epsilon^2} \, ,
\end{equation}
where all variances of observables are with respect to the state $\ket{\mathbf{\gamma}}$ (see Appendix for a brief derivation of the above expression).

While the objective function and its finite-precision estimator $F_{p,\epsilon}(\mathbf{\gamma})$ are the main quantities of interest, the optimization algorithms can be improved by providing additional information. In the next Sections we will explain how the knowledge of the gradient can be exploited to find the values $(\gamma_1,\ldots,\gamma_p)$ that optimize $F_p(\mathbf{\gamma})$ and how to compute its repetition cost.

\vspace{5mm}
\section{Exploiting the gradient of the objective function}
\label{sec:optimization_methods}

Several methods have been developed over the past decades to tackle optimization problems of smooth functions with many parameters, and many good reviews and books are available on the topic (see, for example, references \cite{Nocedal2000,Boyd2004}). While we are not covering the rigorous mathematical background at any level of detail, we would like to provide an intuition of why using gradient information helps. Intuitively, the problem can be recast as the exploration of a multidimensional surface in which we are seeking for the top of the tallest hill. A derivative-free method is like an explorer who has access exclusively to the altitude values of the point she is standing in and of a few of her previous locations. Instead, more advanced methods allows the explorer to give a look at her neighborhood and determine where the uphill direction points.

Optimization methods involving derivatives of the objective function take advantage of the smoothness and differentiability of the objective function itself to predict, solely from the local information provided by the function value and its derivatives in a single point of the parameter space, its behavior in a larger neighborhood. To increase the confidence in the extrapolated model, higher order derivatives may be computed together with the gradient. For optimization efficiency, one has to balance the improved convergence with the cost of evaluating the derivatives. For example, consider that, for a parameter space of dimension $p$, there are $\mathcal{O}(p^k)$ possible $k$-th order derivatives.

The specificity of each optimization method determines how the knowledge of the gradient is exploited. For concreteness, we consider the case of line-search methods that reduce the search from a multidimensional exploration to a sequence of 1D optimizations. From the initial point, one chooses a promising direction and proceeds to the optimization along such line. How to determine the next search direction constitutes the core of the method and several variants are available. The simplest gradient-free choice is to cyclically iterates through a set of directions, one at a time. If the gradient is available, it is possible to choose, at each iteration, the steepest gradient ascent direction, \emph{i.e.} the direction that maximizes the gain in the objective functions for small changes in the parameters. Of course, gathering more information about the local form of the surface leads to better decisions on the direction to take: If one has access to second and higher-order derivatives, their values can guide the creation of a reliable local model and one can search along the direction of the expected maximum. In practice, going beyond the evaluation of the Hessian matrix (\emph{i.e.} of the second derivatives) is costly and does not provide substantial gain.

Empirically, great balance between computational cost and performance is obtained for quasi-Newton methods that rely on the exact calculation of the gradient, but on an approximated form of the Hessian matrix \cite{Nocedal2000}. The BFGS update rule (named after its four authors) is a common choice in which the approximated inverse Hessian matrix is updated from the difference of the objective function and its gradient calculated before and after the last move. To implement this powerful optimization in the context of hybrid algorithms, one needs access solely to $F_p(\mathbf{\gamma})$ and its gradient. In the next Sections we describe how such information is available for an objective function corresponding to the average of multiple quantum measurement outcomes.

\vspace{5mm}
\section{Gradient evaluation and its repetition cost}
\label{sec:gradient_gen}

In analogy with the objective function, also its gradient components are evaluated with finite precision in experiments. Limited precision may be solely caused by the statistical uncertainty associated with quantum observables, or the estimator may be further affected by the use of approximations. The latter case is realized when, for example, finite difference derivatives are employed. Below we describe both the finite difference approach and that based on the analytical expression for the gradient components.

\subsection{Finite difference derivatives}
\label{sec:finite_difference}

Several finite difference schemes exist to approximate the value of derivatives, depending on the required accuracy and the number of function evaluations needed. To quadratic accuracy in the finite increment $\delta$, the formula for the central finite difference derivative is:
\begin{equation}
\label{eq:finite_diff}
	\frac{\partial F_p(\mathbf{\gamma})}{\partial \gamma_n}
    	= \frac{F_p(\gamma_1,\ldots,\gamma_n+\delta/2,\ldots,\gamma_p)-F_p(\gamma_1,\ldots,\gamma_n-\delta/2,\ldots,\gamma_p)}{\delta} + \mathcal{O}(\delta^2) \, .
\end{equation}

Substituting the estimators in the expression at the nominator, one has
\begin{equation}
\label{eq:finite_diff}
	\frac{F_{p,\epsilon^\prime}(\ldots,\gamma_n+\delta/2,\ldots)-F_{p,\epsilon^\prime}(\ldots,\gamma_n-\delta/2,\ldots)}{\delta}
    	= \frac{\partial F_p(\mathbf{\gamma})}{\partial \gamma_n} + \mathcal{O}(\delta^2) + \mathcal{O}(\tfrac{\epsilon^\prime}{\delta}) \, ,
\end{equation}
where $\mathcal{O}(\delta^2)$ is related to the accuracy of the estimate whereas $\mathcal{O}(\tfrac{\epsilon^\prime}{\delta})$ is related to its precision. Notice that we used a precision $\epsilon^\prime$ for the function estimates at the nominator, and it can differ from the one considered in Section~\ref{sec:hybrid_algorithms}.
Why is this flexibility necessary? First of all, it is questionable to require higher precision than accuracy, and this suggests the condition $\epsilon^\prime \geq \delta^3$. Second, $\epsilon^\prime$ should be small enough to separate the values of the estimators at numerator by more than their common precision: $\sqrt{2} \epsilon^\prime \leq \delta \frac{\partial F_{p,\epsilon^\prime}(\mathbf{\gamma})}{\partial \gamma_n}$ , where $\sqrt{2}$ comes from having summed the standard deviations in quadrature.

While the first condition confirms that a high precision is not necessary when $\delta$ is (relatively) large, the second condition may cause problems near local maxima of the objective function. In those regions, each gradient component tends to zero and this imposes such a high precision, meaning very small $\epsilon^\prime$, that it can be achieved only with a large repetition cost, here quantified by:
\begin{equation}
\label{eq:finite_diff_cost}
	M \,
    \geq \, \frac{\mathrm{Var} [\hat{C}]_{\gamma_n+\delta/2} + \mathrm{Var} [\hat{C}]_{\gamma_n-\delta/2}}{2 (\epsilon^\prime)^2} \,
	\approx \, \frac{\mathrm{Var} [\hat{C}]_{\gamma_n} }{(\epsilon^\prime)^2} \, ,
\end{equation}
where the subscript attached to the variance clarifies what is the value of parameter $\gamma_n$ for the corresponding state, and the approximated equivalence holds for small $\delta$.

We verified through numerical simulations that imposing $\epsilon^\prime \leq \frac{\delta}{\sqrt2} \frac{\partial F_{p,\epsilon^\prime}(\mathbf{\gamma})}{\partial \gamma_n}$ causes the increase of several orders of magnitude to the total repetition cost. The advantage of using gradient-based optimization methods is canceled by this cost. To avoid such situation, we introduced a lower bound to $\epsilon^\prime$ directly related to the precision of function estimates: $\epsilon^\prime \geq \tfrac{1}{10} \epsilon$. The constant $\tfrac{1}{10}$ is arbitrary, but has an intuitive meaning: To compute a single gradient component, at most a factor $10^2$ more repetitions are required than those for a single function evaluation.

In our numerical analysis, we determined $\epsilon^\prime$ according to the formula:
\begin{equation}
\label{eq:finite_diff_cost}
	\epsilon^\prime \, = \, \max \bigg\{ \delta^3 , \tfrac{1}{10} \epsilon ,
    	\min \left\{ \epsilon , \frac{\delta}{\sqrt2} \frac{\partial F_p(\mathbf{\gamma})}{\partial \gamma_n} \right\} \bigg\} \, .
\end{equation}
The finite precision and accuracy of the gradient estimation affect the optimization process: We take them into account in our numerical studies, together with the finite precision for the objective function.

\subsection{Analytical gradient}
\label{sec:analytical_gradient}

It is possible to eliminate the bias in the gradient estimation by substituting the approximated formula in Eq.~\eqref{eq:finite_diff} with its analytical expression. Here, we derive the analytical form of the gradient components before discussing its repetition cost. For ease of notation, consider the operator
\begin{equation}
	\W{n}{k}(\mathbf{\gamma}) = U_k(\gamma_k) \ldots U_{n+1}(\gamma_{n+1}) \, U_n(\gamma_n)
\end{equation}
that summarizes the state preparation circuit including only the steps $\{n,n+1,\ldots,k-1,k\}$, with $1\leq n\leq k\leq p$. The operator $\W{1}{p}$ corresponds to the complete state preparation, consistently with the notation in Section~\ref{sec:hybrid_algorithms}. Furthermore, it is helpful to write each gate $\hat{U}_n$ in term of its Hermitian generator $\hat{G}_n$:
\begin{equation}
	\hat{U}_n(\gamma_n) = e^{-i\,\gamma_n\, \hat{G}_n } \, .
\end{equation}
$\hat{G}_n$ is an Hermitian matrix that, similarly to $\hat{C}$, can be expressed as a linear combination of unitaries as $\hat{G}_n =\sum_{\mu}^{k_G} g_\mu\,\sG{\mu}$. Observe that both $k_G$ and $\sG{\mu}$ should depend on the index $n$, but we omit it from the notation for readability. Again, $g_\mu$ can be taken to be real and positive and the number $k_G$ of terms, realistically, polynomial in the number $N$ of qubits. Notice for example that $k_G$ is $\mathcal{O}(1)$ when $\hat{U}_n(\gamma_n)$ corresponds to a one- or two-qubit gate.

The gradient components can be written as:
\begin{align}
\label{eq:grad_component}
	\frac{\partial F_p(\gb)}{\partial \gamma_n} &= i \bra{\phi_0} \, \Wd{1}{n-1} \, \hat{G}_n \, \W{n}{p\,\dagger} \, \hat{C} \, \ket{\mathbf{\gamma}} + \text{h.c.} \nonumber\\
    	&= -2 \, \sum_{\mu=1}^{k_G} \sum_{\nu=1}^{k_C} \, g_\mu \, c_\nu \, \Im \left[ \bra{\mathbf{\gamma}} \, \W{n}{p} \, \sG{\mu} \, \Wd{n}{p} \, \sC{\nu} \, \ket{\mathbf{\gamma}} \right]
\, ,
\end{align}
where h.c. refers to the Hermitian conjugate and $\Im[*]$ denotes the imaginary part. While the first line of the right-hand-side is directly computable with a quantum emulator (\emph{i.e.} in a classical simulation of the quantum system including operations not available in the quantum experiments), expression like $\Wd{1}{n-1} \, \hat{G}_n \, \W{n}{p\,\dagger} \, \hat{C}$ are neither unitary operators nor Hermitian observables. Still, it is possible to evaluate such quantities with quantum circuits based on the second line of Eq.~\eqref{eq:grad_component}.

There is no unique way to construct the estimator for the left hand side of Eq.~\eqref{eq:grad_component}. It depends on the chosen decomposition of $\hat{C}$ and $\hat{G}_n$, but also on the exact way the terms are measured in experiments. Different estimators for the same quantity exhibit different statistical variances, and the repetition cost varies accordingly (in the context of VQE see, for example, reference \cite{McClean2016}). In the rest of this subsection we consider fairly general decompositions, but describe in details the quantum circuit used to measure each term.

\begin{figure*}[t!]
\begin{center}
\includegraphics[width=0.7\linewidth]{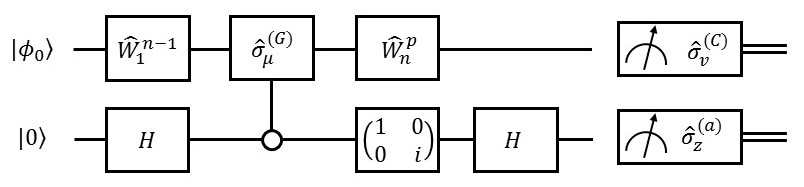}
\caption{Quantum circuit to evaluate the $n$-th component of the gradient in hybrid quantum-classical algorithms, term by term. A single ancilla qubit, labeled $(a)$ and included at the bottom, is required in addition to the $N$-qubit register.The expectation value of the observable $\sC{\nu}\otimes\sz{a}$ with respect to the state $\ket{\psi_\mu}$ at the end of the circuit corresponds to $- \, \Im \left[ \bra{\mathbf{\gamma}} \, \W{n}{p} \, \sG{\mu} \, \Wd{n}{p} \, \sC{\nu} \, \ket{\mathbf{\gamma}} \right]$ as derived in the Appendix.}\label{fig:quantum_circuit_arxiv}
\end{center}
\end{figure*}

In addition to the quantum register with $N$ qubits, we require a single ancilla qubit. Consider the circuit illustrated in Fig.~\ref{fig:quantum_circuit_arxiv}; it generates the final state $\ket{\psi_\mu}$ such that the expectation value of the observable $\sC{\nu}\otimes\sz{a}$ corresponds to
$- \, \Im \left[ \bra{\mathbf{\gamma}} \, \W{n}{p} \, \sG{\mu} \, \Wd{n}{p} \, \sC{\nu} \, \ket{\mathbf{\gamma}} \right]$.
The gradient components can be computed by summing up the expectation values of the different contributions for $\mu\in\{1,2,\ldots,k_G\}$ and $\nu\in\{1,2,\ldots,k_C\}$, with the appropriate weight $c_\nu\,g_\mu$.
While the detailed proof is contained in the Appendix, it is important to make a few observations: The first is that, in case $\hat{C}$ can be directly measured, one can avoid the summation over $\nu$ and directly measure $\hat{C}$. The second is that, in case $\hat{G}_n$ corresponds to an unitary operator, one can avoid the summation over $\mu$ by directly applying gate $\hat{G}_n$. The third is that the role of $\sG{\mu}$ and $\sC{\nu}$ is effectively interchangeable and the proposed quantum circuit can be straightforwardly adapted to the case in which one applies $\sC{\nu}$ on the quantum register and measure $\sG{\mu}$. Analogously, the corresponding summation can be avoided if $\hat{C}$ corresponds to a unitary operator or $\hat{G}_n$ to a directly measurable observable.

In several hybrid variational algorithms one or more of the conditions above are indeed satisfied: In the Variational Quantum Eigensolver (discussed in the introduction) the generators $\hat{G}_n$ are typically products of Pauli matrices and all commuting $\sC{\nu}$ can be measured at the same, while for the Quantum Approximate Optimization Algorithm (analyzed in later Sections) $\hat{C}$ is obtained by measuring the qubit register in the computational basis. For specific variational aglrotithms, this fact greatly simplifies the expression below for the repetition cost and helps to considerably reduce the cost itself.

The repetition cost to achieve the estimate of the $n$-th component of the analytical gradient to a precision $\epsilon^{\dprime}$ is:
\begin{equation}
\label{eq:anal_grad_cost}
	M \,
    \geq \, \frac{ 4 }{(\epsilon^{\dprime})^2}
    	 \, \left\{ \sum_{\mu=1}^{k_G} \sum_{\nu=1}^{k_C} \, g_\mu^2 \, c_\nu^2 \, \mathrm{Var} [\sC{\nu}]_{\mu} \right\} \, ,
\end{equation}
where the variance refers to the observable $\sC{\nu}$ measured on the state $\ket{\psi_\mu}$ produced as output of the quantum circuit in Fig.~\ref{fig:quantum_circuit_arxiv}.

\vspace{5mm}
\section{A representative quantum-classical algorithm: QAOA}

It is difficult to draw conclusions from the above discussion on optimization methods and their repetition cost for the general class of hybrid quantum-classical algorithms based on the variational approach. To analyze the interplay between precision, cost and performance, one needs to focus on specific algorithms. We study the Quantum Approximate Optimization Algorithm (QAOA) recently proposed to solve constraint satisfaction problems. These problems are important in many areas of science and technology but, in their abstract form, involve functions of binary variables: Simply posed, each instance of a constraint satisfaction problem is constituted by a number of clauses that have to be satisfied, and the central question is to determine the maximum number of clauses that can be simultaneously satisfied. An approximated solution is represented by the variable assignment that satisfies a number of clauses corresponding to a high fraction of the best case.

Following an established procedure to solve constrain satisfaction problems with quantum computers, one converts the classical objective function $C(z_1,\ldots, z_n)$, \emph{i.e.} the sum of all the clauses (each close involves a subset of the $N$ binary variables $z_i\in\{-1,1\}$ for $i=1,2,\ldots,N$, and is equal to 1 if satisfied or is otherwise null) into its quantum version. In practice, one substitutes the binary variable $z_i$ with the Z Pauli operator of the $i$-th qubit $\sz{i}$ (more formally, we include in $\sz{i}$ the tensor product of identity operators acting on all other qubits):
\begin{align}
\label{eq:objective_function_QAOA}
	\hat{C}(\sz{1},\ldots,\sz{n}) &= \sum_\alpha \hat{C}_\alpha(\sz{1},\ldots,\sz{n}) \, ,
\end{align}
where the sum over $\alpha$ extends to all the clauses.

The Quantum Approximate Optimization Algorithm specifies how trial states are prepared: One initializes the qubit register in the state $\ket{s}=\ket{++\ldots+}$, corresponding to the balanced superposition of all the $2^N$ bit strings, and then repeatedly applies two quantum operations:
\begin{align}
\label{eq:QAOA_operations}
	\hat{U}(\gamma) &= \exp{(-i \gamma \hat{C})} \nonumber \\
	\hat{V}(\beta)  &= \exp{(-i \beta  \hat{B})} \, ,
\end{align}
with $\hat{B}=\sum_{i=1}^N \sx{i}$. Every time an operation is applied, the parameters $\gamma\in[0,2\pi[$ and $\beta\in[0,\pi[$ can take different values, so that the complete description of a $p$-depth circuit for state preparation is provided by the $2p$-dimensional array $(\gamma_1,\beta_1,\ldots,\gamma_p,\beta_p)=(\gb)$.
Observe that the notation is slightly different from the previous Sections and, in particular, the parameters are now labeled $(\gb)$ instead of $\mathbf{\gamma}$ alone and are in number of $2p$. The specific QAOA reference state is $\ket{\phi_0}=\ket{s}$.

To verify what is the average number of clauses satisfied by the $N$-qubit quantum state $\ket{\gb}$, one needs to compute the expectation value of the observable $\hat{C}$ as defined in \eqref{eq:objective_function_QAOA}. In practice, this is achieved by measuring the qubit register in the computational basis, computing the corresponding value of the objective function $C$, and then averaging over several repetitions of the measurement. The final goal is to find the maximum value of $F_p(\gb)$ as defined by:
\begin{equation}
\label{eq:QAOA_objective_function}
	F_p(\gb) = \bra{\gb} \hat{C} \ket{\gb} \, ,
\end{equation}
over all the accessible states
\begin{equation}
\label{eq:QAOA_state}
	\ket{\gb} = \hat{V}(\beta_p)\,\hat{U}(\gamma_p) \ldots \hat{V}(\beta_1)\,\hat{U}(\gamma_1) \ket{s} \, .
\end{equation}

The advantage of QAOA over other hybrid algorithms is twofold: First, QAOA clearly defines the form of the quantum gates for state preparation and describes them with relatively few parameters \cite{Farhi2014,Farhi2016a}. Second, it has been proven that the exact solution is reachable for state preparation circuits that are deep enough (\textit{i.e.} the limit $p\rightarrow\infty$ of $\max_{(\gb)} F_p(\gb)$ corresponds to the exact global solution \cite{Farhi2014}), in this way certifying the good choice of the state ``ansatz''. It is an open question to address the QAOA performance for small $p$, but there are arguments for a moderate optimism \cite{Farhi2014,Farhi2016a} (see also our numerical results in the next Section). Finally, since for QAOA all terms in $\hat{C}$ involve only $Z$ Pauli operators, one can simply measure every qubit in the computational basis and compute:
\begin{equation}
\label{eq:objective_function_2}
	F_p(\gb) = \sum_{\mathbf{z}\in\{-1,1\}^N} p_\mathbf{Z} \, C(\mathbf{z})
\end{equation}
where the probability $p_\mathbf{Z}=\left|\Braket{\mathbf{z}|\gb}\right|^2$ is estimated through the relative frequency with which the $N$-bit string $\mathbf{z}$ appears among the measurement outcomes. In practice $\hat{C}$ is directly observable in most, if not all, experimental setups.

The formulas for computing the repetition cost to estimate the objective function, the finite difference gradient and the analytical gradient are direct specialization of the expressions in Eq.~\eqref{eq:objective_function_cost}, \eqref{eq:finite_diff_cost} and \eqref{eq:anal_grad_cost} respectively. Furthermore, the analytic form of the gradient components of type-$\gamma$ is:
\begin{align}
\label{eq:g_component}
	\frac{\partial F_p(\gb)}{\partial \gamma_n} &= i \bra{s} \, \Wd{1}{n-1} \, \hat{C} \, \W{n}{p\,\dagger} \, \hat{C} \, \ket{\gb} + \text{h.c.} \nonumber\\
	&= -2 \, \sum_{\mu=1}^{k_C} \, c_\mu \, \Im \left[ \bra{\gb} \, \W{n}{p} \, \sC{\mu} \, \Wd{n}{p} \, \hat{C} \, \ket{\gb} \right]
        \, ,
\end{align}
where we have used the fact that $\hat{C}$ can be directly measured and that the generator $\hat{G}_n$ in this case corresponds to $\hat{C}$, whereas for the components of type-$\beta$ it is:
\begin{align}
\label{eq:b_component}
	\frac{\partial F_p(\gb)}{\partial \beta_n} &= i \bra{s} \, \Wd{1}{n} \, \hat{B} \, \Wd{n+1}{p} \, \hat{C} \, \ket{\gb} + \text{h.c.} \nonumber\\
	&= -2 \, \sum_{i=1}^{N} \, \Im \left[ \bra{\gb} \, \W{n}{p} \, \sx{i} \, \Wd{n}{p} \, \hat{C} \, \ket{\gb} \right]
\, ,
\end{align}
where we have used the explicit decomposition $\hat{B}=\sum_{i=1}^N \sx{i}$.

\vspace{5mm}
\section{Numerical study of QAOA}
\label{sec:numerical_study}

We study the Quantum Approximate Optimization Algorithm as applied to the MAX-CUT problem on random 3-regular graphs. Each instance is defined by an undirected graph and one is asked to color each of the $N$ nodes with one of two colors (consider here the color labels $\{-1,1\}$).  MAX-CUT refers to find the coloring pattern such that the lowest possible number of vertices connect nodes with the same color. According to the notation in terms of binary variables, the vertex between node $i$ and $j$ corresponds to the clause $(1-z_i z_j)/2$. In a 3-regular graphs, each nodes has exactly 3 vertices. Explicitly, and neglecting an additive constant, Eq.~\eqref{eq:objective_function_QAOA} becomes:
\begin{align}
\label{eq:objective_function_MAXCUT}
	\hat{C} &= \sum_{\mu=(\mu_1,\mu_2)} \hat{C}_\mu
    		= -\tfrac{1}{2} \sum_{\mu=(\mu_1,\mu_2)} \sz{\mu_1} \sz{\mu_2} \, ,
\end{align}
in which each vertex $\mu$ is identified by the two nodes it connects, namely $(\mu_1,\mu_s)$. The rightmost side represents the decomposition of $\hat{C}$ that is used to estimate the analytical gradient ($\gamma$-type components) according to $k_C=3N/2$, $c_\mu=1/2$ and $\sC{\mu}=-\sz{\mu_1} \sz{\mu_2}$.

We consider two (classical) optimization methods to identify the values of $(\gb)$ that maximize $F_p(\gb)$: The Nelder-Mead simplex method that does not involve derivatives, and the quasi-Newton method with BFGS update rule (for the approximated inverse Hessian matrix). For the latter optimization we implement either finite difference or analytical gradient.
While both methods empirically demonstrated to be very effective, in presence of a non-convex objective function there is no guarantee to converge towards the global maximum, as opposed to be attracted by a local maximum. For this reason we try to solve each instance starting the optimization process from different, randomly chosen, initial values of $(\gb)$. To summarize the situation, we have:
\begin{itemize}
\item One problem class: MAX-CUT on random 3-regular graphs;
\item Multiple instances: Each characterized by a graph with $N$ nodes and $3N/2$ vertices;
\item Multiple optimization runs for each instances: Characterized by the initial values of $(\gb)$;
\item Three different optimization algorithms: The gradient-free Nelder-Mead (NM) and the quasi-Newton with finite derivative (FD) or analytical gradient (AG).
\item Multiple choices of precision $\epsilon$, and possibly finite increment $\delta$ (FN) or precision $\epsilon^{\dprime}$ (AG).
\end{itemize}
All numerical simulations in this work are performed using qHiPSTER, a state-of-the-art quantum emulator developed by Intel and Harvard University \cite{Smelyanskiy2016a,Sawaya2016a}. The effect of finite precision for all observable quantities is included by adding a stochastic contribution to the expectation value of the single observable (for example, each term of the sum in Eq.~\eqref{eq:g_component}-\eqref{eq:b_component} has a stochastic contribution). The stochastic contribution is drawn uniformly at random in $[-\tilde{\epsilon},\tilde{\epsilon}]$, where $\tilde{\epsilon}$ denotes the precision of the corresponding observable. All data are computed for $N=16$ qubits.

\subsection{Single instance}
\label{sec:single_instance}

We fix the (randomly generated) MAX-CUT instance and analyze the repetition cost and performance of the optimization algorithms. To facilitate the comparison between distinct instances, we define the figure of merit of the ``performance'' not through the maximum $F_p(\gb)$ achieved during the optimization, but via the ratio between the number of satisfied clauses for the approximated solution $\ket{\gb}$ compared to the maximum number of satisfiable clauses (\emph{i.e.} the exact solution):
\begin{equation}
\label{eq:fraction}
	R_p(\gb) = \frac{F_p(\gb)}{ \max_{ \mathbf{z}\in\{-1,1\}^N } C(\mathbf{z})}
\end{equation}

\begin{figure*}[t!]
\begin{center}
\includegraphics[width=0.6\linewidth]{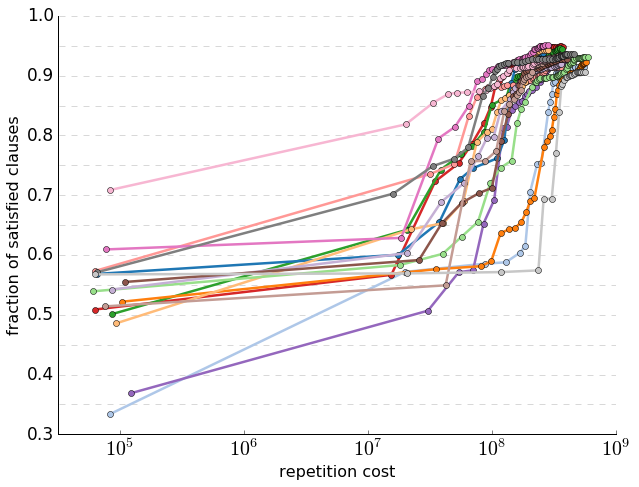}
\caption{Fraction of satisfied clauses $R_7(\gb)$ for circuits of depth $p=7$ \emph{vs.} the repetition cost of the optimization run. Single instance (number 111 in our simulations), $N_r=16$ runs of quasi-Newton method with $\epsilon=0.01$ and $\delta=0.1$.}\label{fig:one_instance_one_method}
\end{center}
\end{figure*}

Fig.~\ref{fig:one_instance_one_method} shows how the initial values of the parameters affect the optimization process. We observe that different optimization runs reach distinct states. In actual applications, one would select the run leading to the approximated solution with highest $R_p(\gb)$ as the one generating the answer from the hybrid algorithm. For the repetition cost it is tempting to consider only that of this post-selected optimization run, but such consideration would be misleading. In absence of a way to \emph{a priori} select good initial values of $(\gb)$, one needs to explore several initial conditions (by contrast, see an alternative approach based on machine learning techniques in reference \cite{Wecker2016}) and, even more, one needs to determine when to stop exploring new initial conditions. In our study we perform $N_r=16$ runs per instance (per optimization method).

\begin{figure*}[t!]
\begin{center}
\includegraphics[width=0.45\linewidth]{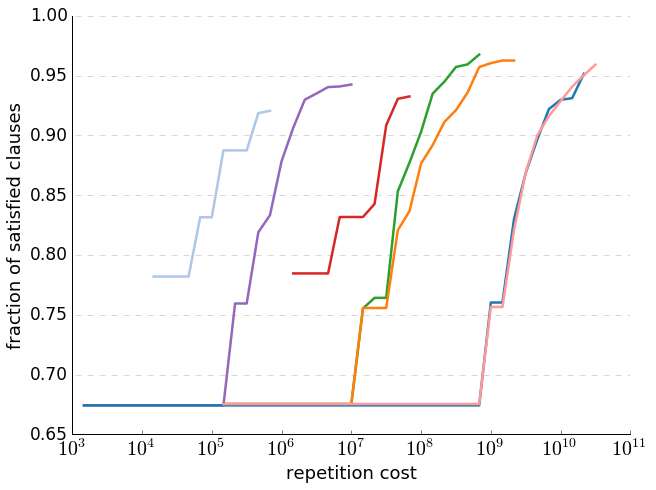}
\hspace{5mm}
\includegraphics[width=0.45\linewidth]{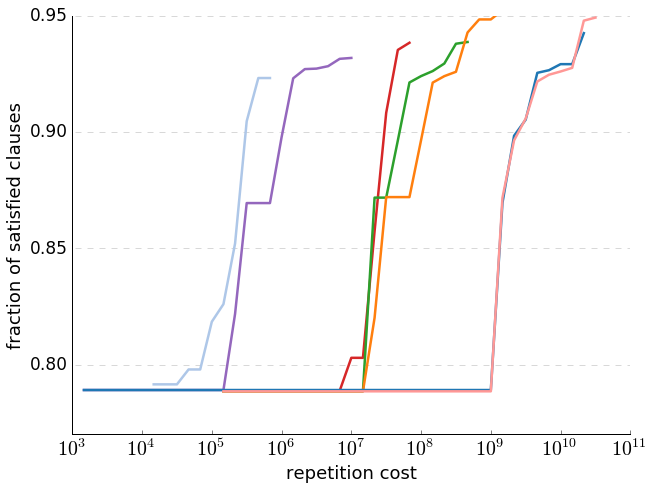}
\includegraphics[width=0.45\linewidth]{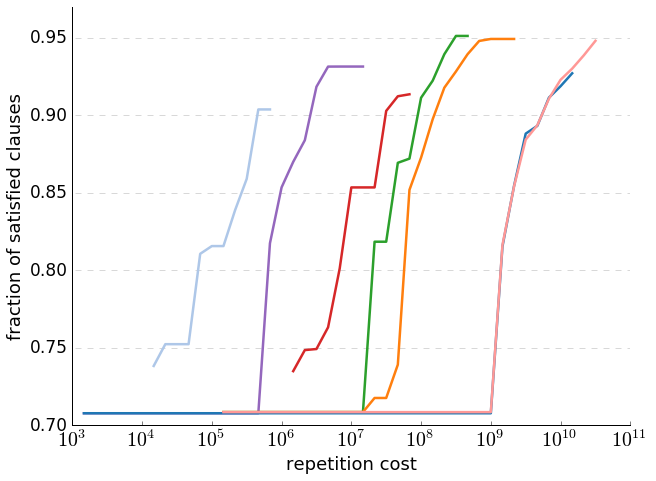}
\hspace{5mm}
\includegraphics[width=0.45\linewidth]{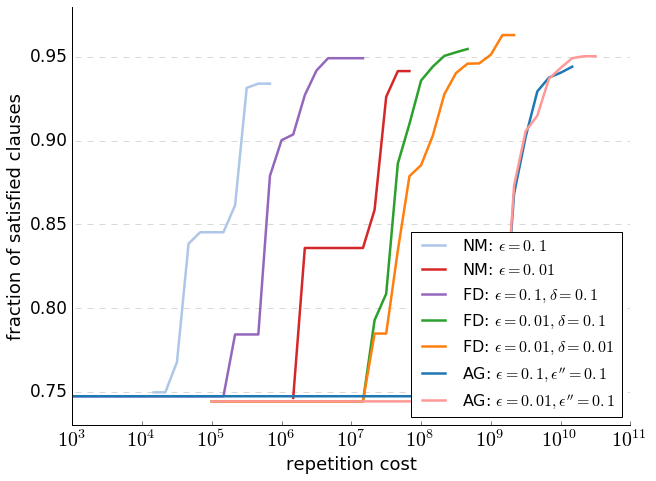}
\caption{Fraction of satisfied clauses $R_7(\gb)$ for circuits of depth $p=7$ \emph{vs.} the repetition cost of the optimization run. Single instance (from top right panel and clockwise order, instance numbers are 33, 63, 120, and 111), multiple optimization methods. For a given method, each line indicates the fraction achieved by the best run as determined at the corresponding repetition cost. For this reason, different parts of the same line may refer to different optimization runs. As explained in the main text, the repetition cost should in general be multiplied by the number of runs, here $N_r=16$ for all methods. The color scheme in the bottom right is the same for all panels.}\label{fig:one_instance_multiple_methods}
\end{center}
\end{figure*}

Fig.~\ref{fig:one_instance_multiple_methods} provides a direct comparison between the post-selected best run for different optimization methods. Due to the relatively small number of optimization runs, it is not guaranteed that we have explored the full potential of each method. On the other side, increasing the number of runs will alter the low-number-of-repetition part of the curves since the probability to start near highly performing states increases. To avoid bias due to more or less favorable initial conditions, we use the same (randomly generated) conditions for all methods. While this is strictly the case for FD and AG, the NM method requires more than one initial value of the parameters (in fact the initial simplex requires $2 p$ points and, for this reason, NM has the advantage of exploring different areas of the parameter space from the very beginning). One of the initial points of the NM coincides with that of the quasi-Newton methods.

While one would expect that increasing the precision $\epsilon$ would improve the performance, this is not always the case. This effect can partially be understood from the analogy with stochastic gradient methods: An inaccurate gradient estimate may cause the escape from a local maximum attraction basin and allow the converge towards a better one. From the inspection of several instances, we can draw three conclusions: a) Gradient-based optimization with finite derivatives (FD) requires, at given accuracy $\epsilon$, more samples than gradient-free optimization (NM), but also achieves higher fraction of satisfied clauses; b) Increasing the increment $\delta$ for the finite-difference derivatives reduces the sampling cost with limited consequences on the optimization effectiveness; c) the increase in cost for the analytical gradient (AG) is not supported by any considerable improvement of the optimization. To confirm the trends stated above,
in the next subsection we perform the statistical analysis of $N_i=128$ instances.

\subsection{Multiple instances}
\label{sec:multiple_instances}

We associate to each instance the largest value of $R_p(\gb)$ obtained during any of the $N_r=16$ runs, and we denote it by $\widetilde{R}_p(\gb)$ without explicitly express in the notation that it is characteristic of a specific instance (and method). The corresponding repetition cost is taken to be the sum of the cost of all $N_r$ optimizations. In Table~\ref{table} we report a few important quantities for the statistical analysis of the distribution of the corresponding fractions of satisfied clauses. Specifically, we report the arithmetic average, standard deviation and median, together with the repetition cost.

\begin{table*}[t!]
\centering
\begin{tabular}{ | l || c | c | c | c |}
\hline
    \hspace{1.1cm} method & average & std. dev. & median & repetition cost \\
\hline
\hline
    NM: $\epsilon=0.1$ & 0.9232 & 0.0194 & 0.9216 & 1.19$\times 10^7$ \\
\hline
    NM: $\epsilon=0.01$ & 0.9291 & 0.0182 &  0.9247 & 1.30$\times 10^9$ \\
\hline
    FD: $\epsilon=0.1$, $\delta=0.1$ & 0.9435 & 0.0171 & 0.9414 & 1.10$\times 10^8$ \\
\hline
    FD: $\epsilon=0.01$, $\delta=0.1$ & 0.9541 & 0.0138 & 0.9517 & 6.68$\times 10^9$ \\
\hline
    FD: $\epsilon=0.01$, $\delta=0.01$ & 0.9539 & 0.0144 & 0.9532 & 2.29$\times 10^{10}$ \\
\hline
    AG: $\epsilon=0.1$, $\epsilon^{\dprime}=0.1$ & 0.9428 & 0.0178 & 0.9423 & 2.24$\times 10^{11}$ \\
\hline
    AG: $\epsilon=0.01$, $\epsilon^{\dprime}=0.1$ & 0.9518 & 0.0147 & 0.9512 & 3.88$\times 10^{11}$ \\
\hline
\end{tabular}
\vspace{2mm}
\caption{Summary table for circuits of depth $p=7$. Average refers to the arithmetic average of the highest fraction of satisfied clauses encountered during each instance, \emph{i.e.} $\widetilde{R}_7(\gb)$. Std. dev. refers to the standard deviation of the distribution over 128 instances. The median of $\widetilde{R}_7(\gb)$ is also provided. The sampling cost includes the fact that several optimization runs (here $N_r=16$) per instance are necessary. See main text for detailed explanations.}\label{table}
\end{table*}

It is important to specify that we implemented specific stopping criteria to ``terminate'' each optimization runs in an automatic way. In fact, if such choice was left arbitrary or based on a case-to-case analysis (for example neglecting a possibly tiny increase in $R_p(\gb)$ provided by a considerable amount of repetitions towards the end of the optimization run), the results related to the repetition cost could be affected by the data post-process that, for its nature, cannot have any effect on the repetition cost for actual experiments (all repetitions have already been performed at that point!). For all practical purposes, we should consider the stopping conditions as an essential part of the method's definition. For details on the stopping criteria we implemented, see Appendix.

\begin{figure*}[t!]
\begin{center}
\includegraphics[width=0.6\linewidth]{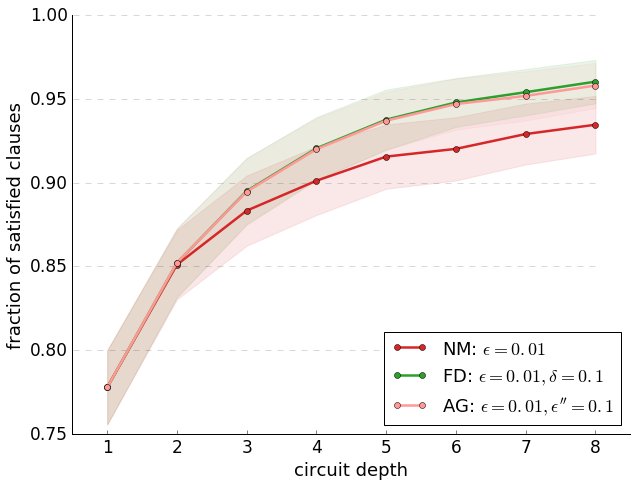}
\caption{Fraction of satisfied clauses $R_p(\gb)$ for circuits of depth $p=1,2,\ldots,8$. The trend Single instance (from top right panel and clockwise order, instance numbers are 63, 33, 7, and 111), best run, multiple optimization methods. For a given method, each line indicates the fraction achieved by the best run as determined at the corresponding repetition cost. For this reason, different parts of the same line may refer to different optimization runs. As explained in the main text, the repetition cost should in general be multiplied by the number $N_r$ of runs, here 16 for all methods.}\label{fig:stat_multiple_methods_multiple_depths}
\end{center}
\end{figure*}

Until this point, we have presented data for a fixed circuit depth, namely for $p=7$. We conducted similar analysis for each circuit depths between $p=1$ and $p=8$. Fig.~\ref{fig:stat_multiple_methods_multiple_depths} shows how the average $\widetilde{R}_p(\gb)$ changes by increasing $p$. Notice that deeper circuits in QAOA are strictly more powerful than shallower ones, but it is also reasonable to expect that the optimization task becomes more challenging. Indeed, this is confirmed by the little difference, if any, at small $p$ where different optimization methods converge to the same maximum value. Quasi-Newton optimization performs consistently better than the Nelder-Mead method while the repetition cost does not differ too much (typically about a factor of 5 in case of finite difference derivative with relatively large increment $\delta=0.1$).

We would like to highlight the importance of future analysis to address the scaling cost of the optimization with respect to the problem size $N$. In this work, as mentioned earlier, we have only studied the case of $N=16$.

To conclude, a final remark on the broad impact of finite precision in the estimation of the objective function: The inexact knowledge of $F_p$ may slow down the optimization process or guide it towards a sub-optimal local maximum, but this is not the only effect. The presence of a stochastic contribution and the fact that we aim for the maximum value of $F_{p,\epsilon}$ contribute to overestimate the value of such local maximum. In our case, since we use a bounded distribution for the stochastic term (drawn uniformly at random in $[-\epsilon,\epsilon]$), the absolute value of such effective bias is at most $\epsilon$. The situation would be different when unbounded distributions, as the Gaussian one, were used to model the finite precision. This aspect and the realistic form of the stochastic term will be the subject of ensuing studies.

\vspace{5mm}
\section{Discussion and conclusions}

Our presentation focused on hybrid algorithms in which the quantum part is based on the gate model of quantum computation. To extend the relevance of our analysis, we observe that quantum-classical variational algorithms have also been introduced in contexts where quantum states evolve in time under the action of a continuously changing Hamiltonian. While in this work we considered a finite set of parameters, the extension to the continuous case requires the optimization of the functional dependence of the parameters. However, a recent study \cite{Yang2016a} showed that the optimal functional dependence is represented by step functions between a small set of values. This reduces the complexity of the optimization since a function can be converted into a set of few parameters (describing, for example, the length of the step interval and its value). Following this approach, gradient-based optimization quite naturally applies to those algorithms too.

In addition, it is important to recognize that our numerical study is limited to the QAOA variant of hybrid variational algorithms. The fact that gradient methods (and quasi-Newton optimizers in particular) seem to be more effective than gradient-free ones does not suffice to claim that such optimization procedures are the most suitable for hybrid schemes in general. First of all, Nelder-Mead is only one of the gradient-free solver developed for optimization problems and some authors observed that in certain contexts it is not the best one of the group \cite{McClean2016}. Second, we do not expect the conclusions of our study to directly extend to cases where $p=\mathcal{O}(10^3)$ or larger, corresponding for example to even moderately large molecule in the UCC+VQE approach.

Finally, note that the presence of noise in the quantum hardware may increase the statistical variance beyond the value due to quantum uncertainty and introduce a bias in the objective function estimator \cite{Sawaya2016a}, and this may only be mitigated, but not eliminated, by the intrinsic robustness of variational algorithms to certain kind of systematic errors \cite{McClean2016}.

In conclusion, hybrid quantum-classical algorithms are likely to play an essential role in the early applications of small- and medium-size quantum devices. However, while ``outsourcing'' part of the computation to classical processing units will likely reduce the overall workload on the quantum hardware, the efficacy of variational algorithms strongly depends on what states can be prepared and on the choice of the optimization method. We investigated the latter aspect and proposed an extension beyond the derivative-free approaches considered so far. Any first order or quasi-Newton method, here we specifically analyzed the BFGS variant, requires the knowledge of the gradient in addition to the objective function. We have presented in detail how to compute the gradient in the general context of hybrid algorithms and for the specific case of the QAOA.

While each optimization method is characterized by several hyper-parameters, for example regulating the balance between the exploration and exploitation phase, and the complete determination of their impact is outside the scope of this work, we explicitly provided the stopping conditions. In fact, if the required precision determines how many repetitions are necessary for a single optimization iteration, the stopping conditions determine when the overall optimization process should terminate and are, therefore, of great experimental relevance.

This is perhaps the most important point, the fact that we have focused the comparison between different variants of the optimization process on metrics relevant to the experimental implementation. In doing so, we included the effect of finite precision and bias (for the finite difference gradient) in our numerical simulations and analyzed the performance in relation to their repetition cost, a metric that reflects the actual feasibility factor in case of shallow quantum circuits.


\section*{Acknowledgments}
The authors would like to thank Nitish Shirish Keskar for discussions on the quasi-Newton optimization, Jhonathan Romero Fontalvo for discussions about gradient methods in VQE and Edward Farhi for discussions on QAOA.



\clearpage

\appendix
\renewcommand\thefigure{\thesection.\arabic{figure}}
\renewcommand\thetable{\thesection.\arabic{table}}
\setcounter{figure}{0}
\setcounter{table}{0}

\section{Repetition cost for quantum variational algorithms}
\label{app:repetition_cost}

\subsection{Repetition cost for the objective function}

The qubit register is prepared in state $\ket{\mathbf{\gamma}}$, one measures the observable $\sC{\nu}$ and obtains the outcome $s_1$. State $\ket{\mathbf{\gamma}}$ is prepared again, measured, and the outcome $s_2$ recorded. The process is repeated for a total of $M_\nu$ times. If $\ket{\mathbf{\gamma}}$ is not an eigenstate of $\sC{\nu}$, the outcomes do not have to coincide. The best estimate of $\bra{\mathbf{\gamma}} \sC{\nu} \ket{\mathbf{\gamma}}$ is given by the arithmetic average of the outcomes $\{s_i\}_{i=1,2,\ldots,M_\nu}$:
\begin{equation}
\label{app:average}
	\bra{\mathbf{\gamma}} \sC{\nu} \ket{\mathbf{\gamma}} \approx \frac{1}{M_\nu} \sum_{i=1}^{M_\nu} s_i \, .
\end{equation}
In the limit $M_\nu\rightarrow \infty$ the expression above is accurate. For a finite number of repetitions, the average of the outcomes belong to a distribution of values centered around the expectation value $\bra{\mathbf{\gamma}} \sC{\nu} \ket{\mathbf{\gamma}}$ and with a variance that is inversely proportional to $N_\nu$. This is formally correct when the outcomes are independent from each other, as is the case for separate experimental repetitions. In Eq.~\eqref{app:average}, the square root of the variance of the right-hand side corresponds to the precision of the estimate of the left-hand side, and is denoted by $\epsilon$:
\begin{equation}
	\epsilon = \sqrt{ \frac{\mathrm{Var} [\sC{\nu}] }{M_\nu} }
    		 = \sqrt{ \frac{\bra{\mathbf{\gamma}} (\sC{\nu})^2 \ket{\mathbf{\gamma}} - \bra{\mathbf{\gamma}} \sC{\nu} \ket{\mathbf{\gamma}}^2 }{M_\nu} } \, .
\end{equation}

When the estimator is the sum of expectation values of multiple observables, for example as in the definition of the objective function $F_p(\mathbf{\gamma}) = \sum_{\nu=1}^{k_C} c_\nu \, \bra{\mathbf{\gamma}} \sC{\nu} \ket{\mathbf{\gamma}}$ in Eq.~\eqref{eq:objective_function_gen} of the main text, then the overall variance of the estimator is:
\begin{align}
	\epsilon^2 = \, \frac{\mathrm{Var} [\hat{C}] }{M} \,
    	      &= \, \frac{\sum_{\nu=1}^{k_C} c_\nu^2 \, \mathrm{Var} [\sC{\nu}]}{M} \nonumber \\
              &= \, \sum_{\nu=1}^{k_C} c_\nu^2 \, \frac{\mathrm{Var} [\sC{\nu}]}{M_\nu} \, ,
\end{align}
where in the second line we have clarified that the total number of outcomes $M=\sum_\nu M_\nu$ is obtained by measuring the different terms $\bra{\mathbf{\gamma}} \sC{\nu} \ket{\mathbf{\gamma}}$ a different number of times. Specifically, the fraction of outcomes for the $\nu$ term is given by:
\begin{equation}
	\frac{M_\nu}{M} \, = k_C \, \frac{c_\nu^2 \mathrm{Var} [\sC{\nu}]}{\sum_{\mu=1}^{k_C} c_\mu^2 \, \mathrm{Var} [\sC{\mu}]} \, ,
\end{equation}
In practice, each term has precision:
\begin{equation}
	\epsilon_\nu = \sqrt{\frac{\mathrm{Var} [\sC{\nu}]}{M_\nu}} = \frac{\epsilon}{\sqrt{k_C}} \, .
\end{equation}

\section{Quantum circuit to estimate the gradient components.}

In Section~\ref{sec:analytical_gradient} of the main text, we introduced a quantum circuit to compute the imaginary part of the overlap $\bra{\mathbf{\gamma}} \, \W{n}{p} \, \sG{\mu} \, \Wd{n}{p} \, \sC{\nu} \, \ket{\mathbf{\gamma}}$. Here we demonstrate that the desired quantity corresponds to the expectation value of the observable $\sC{\nu}\otimes\sz{a}$ over the state $\ket{\psi_\mu}$ at the output of the circuit. To do so, we write the analytical form of the quantum state (quantum register plus single ancilla qubit) at different point of the circuit.

As clear from the graphics in Fig.~\ref{fig:quantum_circuit_arxiv} of the main text, the initial state is $\ket{\phi_0}\otimes\ket{0}$. Before the conditional gate (the only one involving all the qubits), the state is:
\begin{equation}
	\W{1}{n-1}\ket{\phi_0} \otimes \left( \frac{\ket{0}+\ket{1}}{\sqrt2} \right) \, .
\end{equation}
The conditional gate acts like the unitary gate $\sG{\mu}$ on the qubit register if the ancilla qubit is in $\ket{0}$ (the usual convention in quantum information community is to indicate this fact with the empty circle on the ancilla qubit line) and otherwise corresponds to the identity operation. Immediately after we have the state:
\begin{equation}
	\frac{1}{\sqrt2} \left( \W{1}{n-1}\ket{\phi_0}\otimes\ket{0} + \sG{\mu}\W{1}{n-1}\ket{\phi_0}\otimes\ket{1} \right) \, .
\end{equation}

The output of the circuit is:
\begin{align}
	\ket{\psi_\mu} &= \frac{1}{\sqrt2} \left( \W{1}{p}\ket{\phi_0}\otimes\ket{+} + i \W{n}{p}\sG{\mu}\W{1}{n-1}\ket{\phi_0}\otimes\ket{-} \right) \nonumber \\
    			  &= \frac{1}{2} \bigg[ \left( \ket{\mathbf{\gamma}} + i \W{n}{p}\sG{\mu}\W{1}{n-1}\ket{\phi_0} \right) \otimes\ket{0} + \left( \ket{\mathbf{\gamma}} - i \W{n}{p}\sG{\mu}\W{1}{n-1}\ket{\phi_0} \right) \otimes\ket{1} \bigg] \, ,
\end{align}
where in the first line we have used the eigenstates of the X Pauli operator $\ket{\pm}=\tfrac{1}{\sqrt2}(\ket{0}\pm\ket{1})$ and in the second line we expressed $\W{1}{p}\ket{\phi_0}=\ket{\mathbf{\gamma}}$ according to the notation introduced in the main text.

Recalling that $\bra{i}\sz{a}\ket{j}=\delta_{ij}(-1)^j$, by direct computation one can easily verify that:
\begin{equation}
	\bra{\psi_\mu} \sC{\nu}\otimes\sz{a} \ket{\psi_\mu}
    	= - \, \Im \left[ \bra{\mathbf{\gamma}} \, \W{n}{p} \, \sG{\mu} \, \Wd{n}{p} \, \sC{\nu} \, \ket{\mathbf{\gamma}} \right] \, .
\end{equation}

\section{Stopping conditions to terminate the optimization run.}

In Section~\ref{sec:multiple_instances} we highlighted the importance specifying the stopping criteria to terminate the optimization process as as essential part of the algorithm itself or, to be more precise, as an essential part of the concrete application of such algorithm. In fact, the repetition cost is strongly dependent on such choice as we verified in our numerical simulations. Here we summarize the stopping conditions used in the three cases we studied.
\begin{description}
\item[Nelder-Mead (NM)] This is a derivative-free method that evaluates the objective function in $(2p+1)$ points of the parameter space, effectively visualized as the vertices of a simplex. The vertex associated with the lowest objective function is updated according to certain rules \cite{Kelley1999}. Sometimes all vertices but the one associated with the highest objective function are updated.
\begin{itemize}
\item We stop the optimization if the value of the best performing vertex did not change in the latest $2\,p\,\alpha$ iterations. In our simulations, we observed that the optimization proceeds through subsequent plateaus since, once a new best-performing vertex is found, almost all the other needs to be updated sequentially before a novel best-performing vertex is found. Conservatively, we chose $\alpha=20$ for our simulations.
\item To avoid evaluating too many points near the local maximum and being ``deceived'' by a particularly favorable stochastic event (for example one that adds to a suboptimal vertex a contribution $\epsilon$ due to the finite precision making it, in experiments, the new best performing one), we halved $\alpha=10$ when the increment in the objective function was less than $\epsilon/2$.
\item A maximum of 8000 simplex updates was also imposed.
\end{itemize}
\item[BFGS with finite difference derivatives (FD)] This method performs sequential line searches along directions determined by the desire of minimizing a quadratic approximation of the objective function reconstructed from its value and the value of its gradient components.
\begin{itemize}
\item We stopped the optimization when the gradient became too flat, indicating that we reach the immediate neighborhood of the local optimum. Specifically, no more line search directions were considered after the L2 norm of the estimated gradient fell below $\sqrt{2p} \max\{10^3,\delta^2\}$.
\item We also avoided further optimization when the maximum between two consecutive line searches improved by less than $10^{-4}$ and at least $2p$ directions were already explored. Notice that $10^{-4} \leq \epsilon, \epsilon^\prime, \epsilon^{\prime\prime}$ in all our simulations.
\item A maximum of 300 line search directions was also imposed.
\end{itemize}
\item[BFGS with analytical gradient (AG)] Differently from the finite difference approach, the use of analytical gradient does not introduce a bias in the estimation of the objective function derivatives. Finite precision is still related to the number of repetitions performed. The stopping criteria were the same as for the finite difference approach FD.
\end{description}





\bibliographystyle{apsrev4-1}
\bibliography{QAOA_gradient}

\end{document}